# PetFMM—A dynamically load-balancing parallel fast multipole library

Felipe A. Cruz [a]  Matthew G. Knepley [b]  L. A. Barba [c],*

[a] *Department of Mathematics, University of Bristol, England BS8 1TW*

[b] *Computation Institute, Argonne National Laboratory and University of Chicago*

[c] *Department of Mechanical Engineering, Boston University, Boston MA 02215*

## Abstract

Fast algorithms for the computation of $N$-body problems can be broadly classified into mesh-based interpolation methods, and hierarchical or multiresolution methods. To this last class belongs the well-known fast multipole method (FMM), which offers $\mathcal{O}(N)$ complexity. The FMM is a complex algorithm, and the programming difficulty associated with it has arguably diminished its impact, being a barrier for adoption. This paper presents an extensible parallel library for $N$-body interactions utilizing the FMM algorithm, built on the framework of PETSc. A prominent feature of this library is that it is designed to be extensible, with a view to unifying efforts involving many algorithms based on the same principles as the FMM and enabling easy development of scientific application codes. The paper also details an exhaustive model for the computation of tree-based $N$-body algorithms in parallel, including both work estimates and communications estimates. With this model, we are able to implement a method to provide automatic, *a priori* load balancing of the parallel execution, achieving optimal distribution of the computational work among processors and minimal inter-processor communications. Using a client application that performs the calculation of velocity induced by $N$ vortex particles, ample verification and testing of the library was performed. Strong scaling results are presented with close to a million particles in up to 64 processors, including both speedup and parallel efficiency. The largest problem size that has been run with the PetFMM library at this point was 64 million particles in 64 processors. The library is currently able to achieve over 90% parallel efficiency for 32 processors, and over 85% parallel efficiency for 64 processors. The software library is open source under the PETSc license, even less restrictive than the BSD license; this guarantees the maximum impact to the scientific community and encourages peer-based collaboration for the extensions and applications.

# 1 Introduction

So-called $N$-body problems arise in many areas (*e.g.* astrophysics, molecular dynamics, vortex methods, electrostatics). In these problems, the system is described by a set of $N$ particles, and the dynamics of the system is the result of the interactions that occur for every pair of particles. To calculate all such interactions, the total number of operations required normally scales as $N^2$. One useful way to mathematically express an $N$-body problem is by means of a matrix-vector multiplication, where the matrix is dense and represents the particle interactions, and the vector corresponds to the weights of the particles. Thus, the mat-vec product corresponds to the evaluation of all pairwise interactions. In a naive implementation, we would directly create the matrix in computer memory and then perform the multiplication with the vector. This naive approach would prove feasible only for small $N$, as the computational requirements in processing power and memory both grow as $N^2$. For this reason, many efforts have been directed at producing efficient implementations, capable of performing the mat-vec operation at reduced memory requirements and operation counts.

The various methods for more efficient calculation of the particle interaction problem can be broadly classified in two types: mesh-based interpolation methods, and hierarchical or multiresolution methods. The basic mesh-based method is the particle-mesh (PM) approach, in which particle information is interpolated onto a lattice, the field of interest is solved on the mesh by means of some grid-based method such as finite difference, and the field information is finally interpolated back to the particle locations [13]. (This method is also called particle-in-cell, PIC.) In some applications, such as molecular dynamics, the smoothing at the short-range introduced by the interpolations in the PM method is unacceptable. An alternative mesh-based method can then be used in which all near-field potentials are calculated directly, while far-field effects are calculated with the PM method; this is called the particle-particle/particle-mesh method (P³M).

The second class of methods provides efficient computation of the interactions by means of a hierarchical or multilevel approach. The main subset of this class performs a hierarchical subdivision of the computational domain, which is used to group particles at different length scales, and then approximates the interactions of clusters of particles using series expansions. The approximation is applied to far-field interactions, while near-field interactions are summed directly. This type of methods can be considered meshfree, and includes: tree codes [5,2], the fast multipole method (FMM) [11] and its variations [7,1,23,9]. An alternative multilevel method has been presented by Skeel [21], which instead of using series expansions for the far field approximations utilizes a multigrid approach. The approximation of the field is in this case performed after a splitting of the particle potentials into a smooth part (matching the far field) and a nonsmooth part (acting only in the near field). While the nonsmooth (purely local) part can be calculated directly at low cost, a multigrid method is used to approximate the smooth component. For this approximation, the method relies on gridded basis functions of compact support, like in P³M, but unlike P³M it provides acceleration via multiple grid levels, achieving $\mathcal{O}(N)$ complexity. Thus, this method applies the multilevel approach via field decomposition, rather than spatial decomposition; it can perhaps be viewed as a hybrid of the mesh-based and hierarchical methods.

* Correspondence to: labarba@bu.edu

In this paper, we present the theory and development of a parallel fast multipole library, PetFMM [1], belonging to the *meshfree* type of methods described above. The overarching goal is to unify the efforts in the development of FMM-like methods into an open-source library, that provides a framework with the capacity to accommodate memory efficiency, parallelism and the data structures used for spatial decomposition. The software is implemented utilizing PETSc, the parallel library for scientific computing developed over more than 17 years at Argonne National Laboratory [3]. At this point, we have a complete implementation of the FMM in parallel, with dynamic load balancing provided by means of an optimization approach—minimizing inter-node communications and per-node computational work. But a prominent feature of this implementation is that it is designed to be *extensible*, so that it can effectively unify efforts involving many algorithms which are based on the same principles of the FMM. The perspectives for extensibility are described in this paper as well.

The development of this extensible parallel library for $N$-body interactions is important due to the programming difficulty associated with the FMM algorithm, which has been a barrier for adoption and arguably diminished its potential impact. A critical stage in the maturation of a computational field begins with the widespread availability of community software for the central algorithms. A good example is molecular dynamics, which blossomed after introduction of freely available packages, such as CHARMM [6] and NAMD [18]. Such game-changing community software does not exist for particle methods or for $N$-body computations in general.

The present paper does not merely describe a new parallel strategy and implementation, it also details an exhaustive model for the computation of tree-based $N$-body algorithms in parallel. Our model is a significant extension of the time model developed in [10], which assumed a uniform distribution of the particles among processors and does not address load balancing or communication overheads. With our model, which includes both work estimates and communication estimates, we are able to implement a method to provide a priori, automatic load balancing.

---

[1] PetFMM stands for 'portable extensible toolkit for FMM', as in PETSc, which is 'portable extensible toolkit for scientific computing'.

The first parallel implementations of the FMM was that of Greengard and Gropp [10], on a shared memory computer. They also presented a timing model for a perfectly balanced implementation, of which we say more in §5. More recent versions of the parallel FMM have been presented in [24,12,17,16]. Many of these codes produce a partition of the data among processors based upon a space-filling curve, as previously introduced for the case of parallel treecodes in [22]. Only one of these parallel FMM codes represents a supported, open source code available for community use. The KIFMM3d code can be downloaded and modified under the GPL license. It is not an implementation of the classic FMM algorithm, however, but rather a kernel-independent version developed by the authors. This algorithm does not utilize series expansions, but instead uses values on a bounding surface of each subdomain obtained through an iterative solution method; these ideas are based on [1]. The code does not appear to allow easy extension to traditional FMM, or to the other related algorithms alluded to above. Moreover, it does not appear to be designed as an embedded library component, part of a larger multiphysics simulation.

This paper is organized as follows. We present first a brief overview of the FMM algorithm; this presentation is necessarily cursory, as a large body of literature has been written about this method. A basic description, however, is necessary to agree on a common terminology for the rest of the paper. Our approach is to illustrate the method using graphical representations. The following section (§3) describes our client application code, the vortex particle method for simulation of incompressible flow at high Reynolds numbers. In this method, the FMM is one of two approaches commonly used to obtain the velocity of particles from the vorticity field information; the second approach (as in other $N$-body problems) is to interpolate information back and forth from a mesh, while solving for the field of interest on the mesh. Although equally efficient—both can be $\mathcal{O}(N)$—the extensive use of interpolations may introduce numerical diffusion, which is undesirable in certain applications. Next, §4 discusses our parallelization strategy. The goal is achieving optimal distribution of the computational work among processors and minimal communication requirement. Our approach to parallelization is original in the use of an optimization method to obtain the parallel partitioning. To be able to apply such an optimization, there is need for good estimates of the computational work required for algorithmic components, as well as communication requirements. The development of these estimates, in addition to memory estimates, is presented in §5. The subsequent section (§6) presents details of our software design, implementation and verification carried out. Results of computational experiments with the parallel software are presented in §7, and we end with some conclusions and remarks about future work.

Note that a very simple and flexible, yet efficient and scalable code has been produced. The entire PetFMM implementation of FMM is only 2600 lines of C$^{++}$, including comments and blank lines. It is already freely downloadable from **http://petsc.cs.iit.edu/petsc/** and we welcome correspondence with potential users or those who wish to extend it for specific purposes.

## 2 Overview of the Fast Multipole Method algorithm

The fast multipole method (FMM) is an algorithm which accelerates computations of the form:

$$f(y_j) = \sum_{i=1}^{N} c_i \, \mathbb{K}(y_j, x_i) \tag{1}$$

representing a field value evaluated at point $y_j$, where the field is generated by the influence of sources located at the set of centers $\{x_i\}$. The sources are often associated with particle-type objects, such as stellar masses, or charged particles. The evaluation of the field at the centers themselves, therefore, represents the well-known $N$-body problem. In summary: $\{y_j\}$ is a set of evaluation points, $\{x_i\}$ is a set of source points with weights given by $c_i$, and $\mathbb{K}(y, x)$ is the kernel that governs the interactions between evaluation and source particles. The objective is to obtain the field $f$ at all the evaluation points, which requires in principle $\mathcal{O}(N^2)$ operations if both sets of points have $N$ elements. Fast algorithms aim at obtaining $f$ approximately with a reduced operation count, ideally $\mathcal{O}(N)$.

The FMM works by approximating the influence of a cluster of particles by a single collective representation, under the assumptions that the influence of particles becomes weaker as the evaluation point is further away, *i.e.*, the kernel $\mathbb{K}(y, x)$ decays as $|x - y|$ increases, and that the approximations are used to evaluate far distance interactions. To accomplish this, the FMM hierarchically decomposes the computational domain and then it represents the influence of sets of particles by a single approximate value. The hierarchical decomposition breaks up the domain at increasing levels of refinement, and for each level it identifies a near and far sub-domain. By using the hierarchical decomposition, the far field can be reconstructed as shown in Figure 1.

Using the computational domain decomposition, the sum in Equation (1) is decomposed as

$$f(y_j) = \sum_{l=1}^{N_{near}} c_l \mathbb{K}(y_j, x_l) + \sum_{k=1}^{N_{far}} c_k \mathbb{K}(y_j, x_k) \tag{2}$$

where the right-most sum of (2), representing the far field, is evaluated approximately and efficiently.

We now need to introduce the following terminology with respect to the mathematical tools used to agglomerate the influence of clusters of particles:

**Multipole Expansion (ME):** a series expansion truncated after $p$ terms which represents the influence of a cluster of particles, and is valid at distances large with respect to the cluster radius.

**Local Expansion (LE):** a truncated series expansion, valid only inside a sub-domain, which is used to efficiently evaluate a group of MEs.

In other words, the MEs and LEs are series (*e.g*, Taylor series) that converge in different sub-domains of space. The center of the series for an ME is the center of the cluster of source particles, and it only converges outside the cluster of particles. In the case of an LE, the series is centered near an evaluation point and converges locally.

As an example, consider a particle interaction problem with decaying kernels, where a cluster of particles far away from an evaluation point is 'seen' at the evaluation point as a 'pseudo-particle', and thus its influence can be represented by a single expression. For example, the gravitational potential of a galaxy far away can be expressed by a single quantity locally. Thus, by using the ME representing a cluster, the influence of that cluster can be rapidly evaluated at a point located far away —as only the single influence of the ME needs to be evaluated, instead of the multiple influences of all the particles in the cluster. Moreover, for clusters of particles that are farther from the evaluation point, the pseudo-particle representing that cluster can be larger. This idea, illustrated in Figure 1(b), permits increased efficiency in the computation.

The introduction of an aggregated representation of a cluster of particles, via the multipole expansion, effectively permits a decoupling of the influence of the source particles from the evaluation points. This is a key idea, resulting in the factorization of the computations of MEs that are centered at the same point, so that the kernel can be written as,

$$\mathbb{K}(x_i, y_j) = \sum_{m=0}^{p} a_m(x_i) f_m(y_j) \tag{3}$$

This factorization allows pre-computation of terms that can be re-used many times, thus increasing the efficiency of the overall computation. Similarly, the local expansion is used to decouple the influence of an ME from the evaluation points. A group of MEs can be factorized into a single LE so that one single evaluation can be used at multiple points locally. By representing MEs as LEs one can efficiently evaluate a group of clusters in a group of evaluation points.

## 2.1 Hierarchical space decomposition

In order to utilize the tools of MEs and LEs, a spatial decomposition scheme needs to be provided. In other words, for a complete set of particles, we need to find the clusters that will be used in conjunction with the MEs to approximate the far field, and the sub-domains where the LEs are going to be used to efficiently evaluate groups of MEs. This spatial decomposition is accomplished by a hierarchical subdivision of space associated to a tree structure (*quadtree* structure in two dimensions, or an *octree* structure in three dimensions) to represent each subdivision. The nodes of the tree structure are used to define the spatial decomposition, and different scales are obtained by looking at different levels of the tree. A tree representation of the space decomposition allows us to express the decomposition independently of the number of dimensions of the space. Consider Figure 1(a) where a quadtree decomposition of the space is illustrated. The nodes of the tree at any given level cover the whole domain. The relations between nodes of the tree, represent spatial refinement. The domain covered by a parent box is further decomposed into smaller sub-domains by its child nodes. Thus, in the FMM the tree structure is used to hierarchically decompose the space and the hierarchical space decomposition is used to represent the near-field and far-field domains. As an example, consider Figure 1(b) where the near-field for the *black* colored box is represented by the light colored boxes, and the far-field is composed by the dark colored boxes.

## 2.2 Bird's eye view of the complete algorithm

After the spatial decomposition stage, the FMM can be roughly summarized in three stages: *upward sweep*, *downward sweep*, and *evaluation step*. In the *upward sweep*, the objective is to build the MEs for each node of the tree. The MEs are built first at the deepest level, level $L$, and then translated to the center of the parent nodes. Thus, the MEs at the higher levels do not have to be computed from the particles, they are computed from the MEs of the child nodes. In the *downward sweep* of the tree, the MEs are translated into LEs for all the boxes in the *interaction list*. At each level, the interaction list corresponds to the cells of the same level that are in the far field for a given cell. Once the MEs have been translated into LEs, the LEs of upper levels are translated and added up to obtain the complete far domain influence for each box at the leaf level of the tree. At the end of the *downward sweep*, each box will have an LE that represents the complete far-field for the box. Finally, at the *evaluation step*, for every particle at every node at the deepest level of the tree, the final field is evaluated by adding the near-field and far-field contributions. The near field of the particles at a given box is obtained by directly computing the interaction between all the particles in the near domain of the box. The far field of the particles is obtained by evaluating the LE of the box at each particle location.

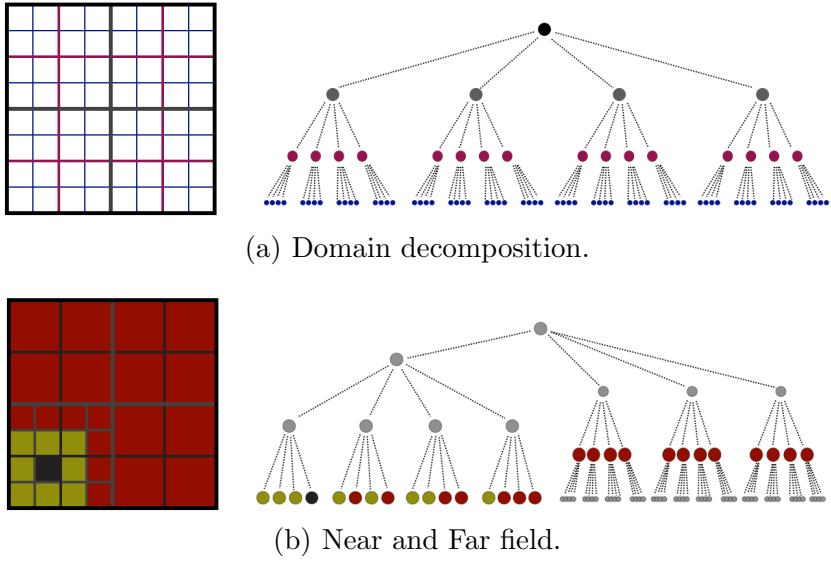

(a) Domain decomposition.

(b) Near and Far field.

Fig. 1. Quadtree decomposition of a two-dimensional domain: (a) presents the hierarchical tree related to the full spatial decomposition of the domain; (b) presents a colored two dimensional spatial decomposition for the black box and its equivalence on the tree. The near-field is composed by the bright boxes and the black box itself, while the far-field is composed by the dark colored boxes. Notice that the far-field is composed of boxes of different levels of the tree structures. The relations between the nodes of the tree simplify the process of composing the near and far domains.

These ideas can be visualized with an illustration, as shown in Figure 2, which we call the "bird's eye view" of the complete algorithm. The importance of this bird's eye view is that it relates the algorithm computations to the data structure used by the fmm. This will prove to be very useful when we discuss the parallel version that we have developed.

This overview is only intended to introduce the main components of the fmm algorithm, so that we can discuss in the forthcoming sections our strategy for parallelization. Therefore, it is not a detailed presentation (we have left out all the mathematics), and readers interested in understanding all the details should consult the original reference [11], and selections of the abundant literature published since.

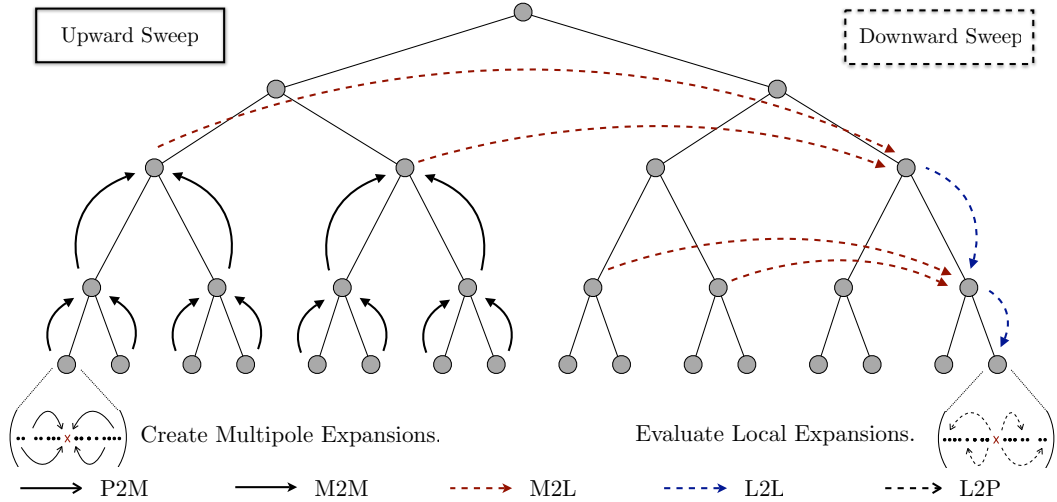

Fig. 2. Bird's eye view of the FMM algorithm. The sketch illustrates the *upward sweep* and the *downward sweep* stages on the tree. Each stage has been further decomposed into the following substages: *P2M*–transformation of particles into MEs (particle-to-multipole); *M2M*–translation of MEs (multipole-to-multipole); *M2L*–transformation of an ME into an LE (multipole-to-local); *L2L*–translation of an LE (local-to-local); *L2P*–evaluation of a LEs at particle locations (local-to–particle).

## 3 Example client application: particle methods

The fast multipole method has many applications, such as calculation of the gravitational field of many stellar bodies, or the electrostatic forces obtained from many charged particles. In fluid dynamics, one application is found in the calculation of the velocity induced by many vortex particles, which in turn is used as a method of simulation for either the Euler or Navier-Stokes equations. The vortex particle method starts by discretizing the continuous vorticity field, defined as the curl of the velocity ($\omega = \nabla \times u$), over a set of moving nodes located at $x_i$ as follows:

$$\omega(x,t) \approx \omega_\sigma(x,t) = \sum_i^N \gamma_i \zeta_\sigma(x,x_i) \tag{4}$$

where a common choice for the basis function is a normalized Gaussian such as:

$$\zeta_\sigma(x,y) = \frac{1}{2\pi\sigma^2} \exp\left(\frac{-|x-y|^2}{2\sigma^2}\right) \tag{5}$$

The discretized vorticity field in (4) is used in conjunction with the vorticity transport equation. For ideal flows in two dimensions, the vorticity equation simply expresses that the vorticity is a preserved quantity over particle tra-

jectories:

$$\frac{\partial \omega}{\partial t} + u \cdot \nabla \omega = \frac{\mathrm{D}\omega}{\mathrm{D}t} = 0 \qquad (6)$$

Therefore, the moving nodes can translate with their local velocity and carry their vorticity value to automatically satisfy the transport equation. The only missing ingredient is obtaining the velocity from the discretized vorticity, which is accomplished using the Biot-Savart law of vorticity dynamics:

$$u(x,t) = \int (\nabla \times \mathbb{G})(x-x')\omega(x',t)dx' = \int \mathbb{K}(x-x')\omega(x',t)dx' = (\mathbb{K} * \omega)(x,t)$$

where $\mathbb{K} = \nabla \times \mathbb{G}$ is the Biot-Savart kernel, with $\mathbb{G}$ the Green's function for the Poisson equation, and $*$ representing convolution. For example, in 2D the Biot-Savart law is written explicitly as,

$$u(x,t) = \frac{-1}{2\pi} \int \frac{(x-x') \times \omega(x',t)\hat{\mathbf{k}}}{|x-x'|^2} dx'. \qquad (7)$$

When the vorticity is expressed as a radial basis function expansion, one can always find an analytic integral for the Biot-Savart velocity, resulting in an expression for the velocity at each node which is a sum over all particles. Using the Gaussian basis function (5), we have:

$$\mathbb{K}_\sigma(x) = \frac{1}{2\pi|x|^2}(-x_2, x_1)\left(1 - \exp\left(-\frac{|x|^2}{2\sigma^2}\right)\right). \qquad (8)$$

where $|x|^2 = x_1^2 + x_2^2$. Thus, the formula for the discrete Biot-Savart law in two dimensions gives the velocity as follows,

$$u_\sigma(x,t) = \sum_{j=1}^N \gamma_j \; \mathbb{K}_\sigma(x-x_j). \qquad (9)$$

Therefore, the calculation of the velocity of $N$ vortex particles is an $N$-body problem, where the kernel $\mathbb{K}(x)$ decays with distance, which makes it a candidate for acceleration using the FMM. Also note that as $|x|$ becomes large, the kernel $\mathbb{K}(x)$ approaches $1/|x|^2$. We take advantage of this fact to use the multipole expansions of the $1/|x|^2$ kernel as an approximation, while the near-field direct interactions are obtained with the exact kernel $\mathbb{K}$. In a separate work, we have investigated the errors of the FMM for the vortex method, and have demonstrated that using the expansions for $1/|x|^2$ does not impact on accuracy, as long as the local interaction boxes are not too small [8].

## 4   Parallelization strategy

The goal of our parallel strategy is to achieve an optimal distribution of the computational work among processors and a minimal communication requirement. We achieve this goal by means of a two-step process: in the first step, we decompose the complete FMM into basic *algorithmic elements*; in the second step, we distribute those algorithmic elements across the available processing elements (processors) in an optimal way. By optimal, we mean that when distributing the basic algorithmic elements we optimize their assignment to the processing elements by:

(1) Load balance: The work performed by each processing element is adequate to the processor's capabilities.
(2) Minimization of communications: The total amount of communication between different processing elements is minimized.

An important ingredient is that both steps of the above-mentioned process are carried out automatically by an optimization tool, without intervention of the user, with the objective of using the available resources with maximum efficiency. The load balancing provided by our method is applied dynamically, *a priori* of the actual computation.

We are aware of one previous attempt to provide automatic load balancing in an FMM implementation. The DPMTA code[19] (not maintained or supported since 2002) is a parallel implementation of the FMM algorithm aimed at molecular dynamics applications. In this code, an initial uniform partitioning of the data at the deepest level of the tree is carried out. Then, elapsed wall clock times are obtained for each of the processors for the equal partitioning, and a data migration scheme is applied to counter load imbalance. Experiments were reported in [20] for $N$=200,000 in 12 processors, showing for the original equal partition elapsed wall clock times for each processor between 60 and 140 seconds. After load balancing, elapsed times were between 80 and 100 seconds, indicating a significant improvement. This approach, however, requires multiple calculations with the same set of particles; it does not seem to be possible to use this approach in an ever-evolving particle configuration, where each configuration is calculated only once. There was also no theory or computational model of work and communications provided in relation to this code. The experiments cited, however, do provide clear evidence that a straightforward uniform data partition (accomplished using a space-filling curve indexing scheme) can result in considerable load imbalance.

In our implementation, we utilize the tree structure associated with the hierarchical decomposition of the domain in order to decompose the FMM into basic algorithmic elements. The tree structure has many roles: it is used as a space partitioner for the particles, it organizes the storage for the multipole expansions and local expansions, and it indicates the relations between nodes in the same level of the tree (if two nodes are from the local domain list or the interaction list). But, most importantly, it can be used to represent the complete algorithm, as mentioned in §2.2. An important part of our approach is to look at the tree not as a data structure, but as a complete description of the *algorithm*; one can see that in Figure 2 all parts of the algorithm are represented. This view is the basis for our parallelization strategy, as described below.

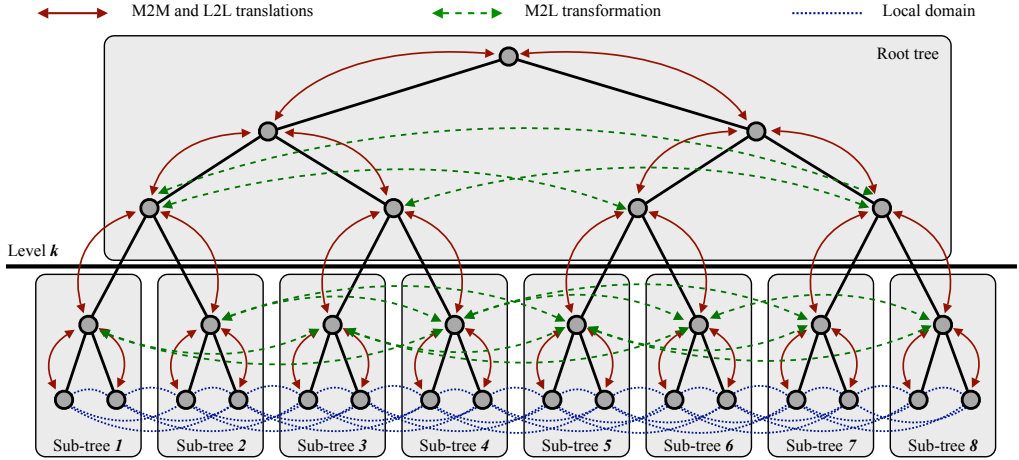

Fig. 3. Illustration of the data usage patterns and partitioning of a binary-tree. In this figure, the tree has been *cut* at level $k = 3$. All data usage patterns that take place between nodes of the tree are illustrated by arrows. The subtrees generated after the partition of the tree are represented by the boxes. Communication between subtrees are generated by data usage patterns that take place between two different subtrees.

The sub-division of the whole algorithm occurs by "cutting" the $d$-dimensional tree at a certain level $k$, as shown in Figure 3. This procedure produces a *root* tree, that contains the first $k$ levels of the original tree, and $2^{dk}$ *local* trees, each corresponding to one of the lower branches of the original tree. The objective of this partitioning strategy is to obtain more subtrees than the number of available processes, so that the subtrees can be optimally distributed across the processes.

In the bird's eye view of the whole algorithm, Figure 2, data and computations are related to nodes of the tree structure. When partitioning the tree representation into subtrees, computations that require data from different nodes of the partitioned tree might access data from several different partitions. If this is the case, communication between partitions will happen as illustrated in Figure 3. By relating computations to nodes of the tree, the work carried out by each partition and the communication between different partitions can be estimated, which is then used to optimally distribute the partitions over available processors.

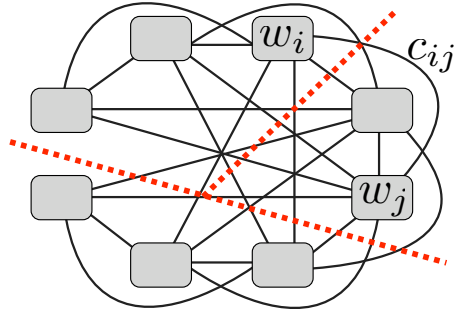

Fig. 4. Sketch of an undirected graph representation of the original hierarchical tree. The $w_i$ weights represent the work performed by the subtree $i$ and the $c_{ij}$ weights represent the communication between subtrees $i$ and $j$. The graph is then partitioned into a number of parts equal to the available processes; the red lines in the figure show a possible partitioning of the graph. The partitions are optimized so that the weights $w_i$ assigned to a partition are balanced with respect to other partitions, and the cost of the cut edges is minimal.

In order to assign subtrees to processors, we build a *graph representation* from the partitioned tree, as illustrated on Figure 4. The graph is assembled such that the vertices of the graph correspond to the subtrees and the edges correspond to communication between subtrees. Using the graph representation, we can assign weights to the vertices which are proportional to the amount of computational work performed by each subtree, and assign weights to the edges which are proportional to the amount of communication that happens between two subtrees.

The load balancing in the parallel algorithm is done by partitioning the weighted graph into as many parts as the number of available processors, and then assigning the subtrees to the processors according to the graph partitions. The problem of obtaining partitions, such that they are well-balanced and minimize the communication, is solved by a graph partitioning tool such as ParMETIS [14]. ParMETIS is an open source graph partitioning tool, and is used by PetFMM to create near optimal partitions. Figure 5 demonstrates the load balancing scheme at work. In an example computation, particles have been placed inside a square-shaped domain that is then hierarchically decomposed into a tree representation. The tree is then cut at level $k = 4$, resulting in 256 parallel subtrees that are then distributed among 16 processors. In the next section we develop the model of the parallel algorithm, obtaining the necessary weights for computational work and communication.

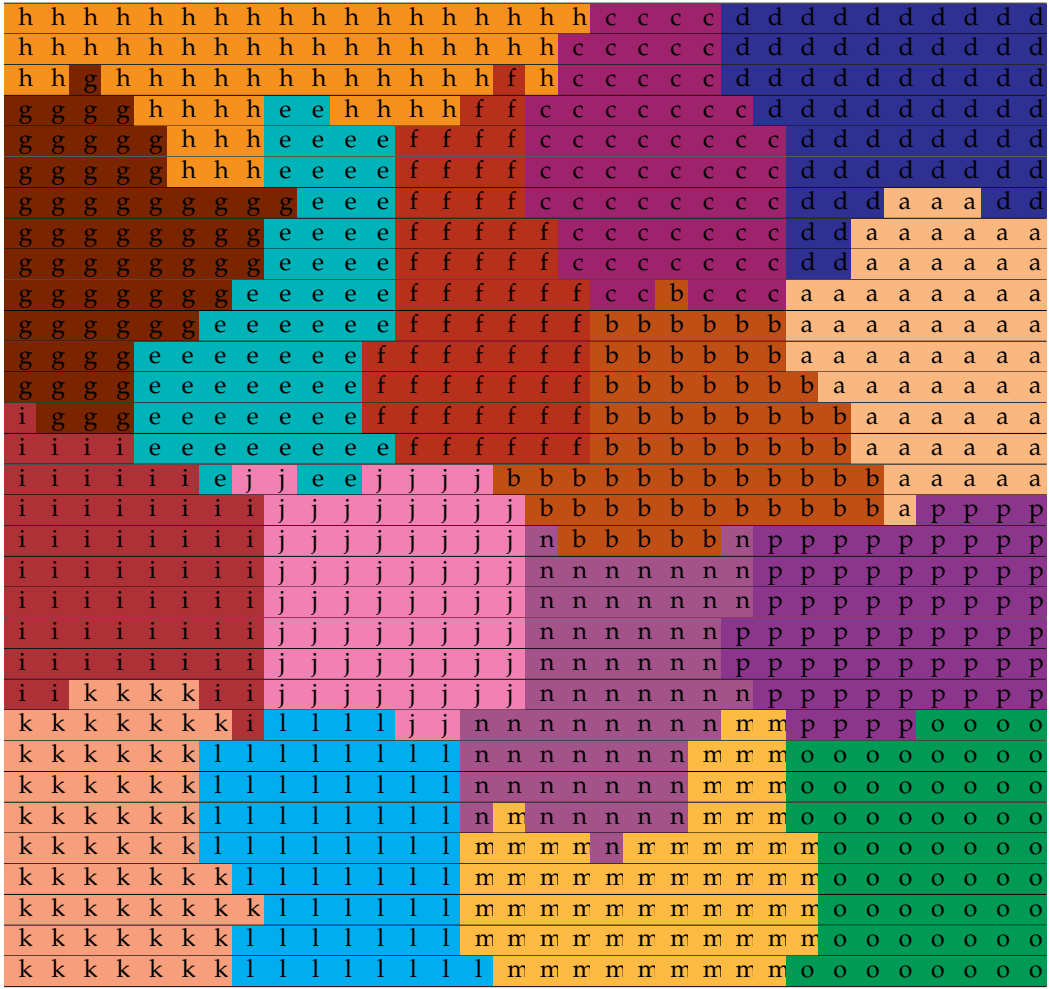

Fig. 5. Illustration of the outcome of the automatic load balancing for a uniform spatial distribution of particles in a square domain. The cells represent the subtrees obtained after the partitioning of the FMM tree. The cells have been labeled and colored to represent the partition to which they have been assigned after the load balancing. In total, 256 subtrees have been distributed among 16 partitions.

## 5  Estimates of work, communication, and memory

The seminal analysis of the parallel complexity of the FMM was given by Greengard and Gropp in 1990 [10]. They develop a model for the running time $T$ in terms on the number of particles $N$, the number of processors $P$, and the number of boxes at the finest level $B$,

$$T = a\frac{N}{P} + b\log_4 P + c\frac{N}{BP} + d\frac{NB}{P} + e(N, P). \qquad (10)$$

They account here for perfectly parallel work, represented by $a$, such as the multipole expansion initialization and local expansion evaluation; the reduc-tion bottleneck, $b$, such as that associated to multipole-to-multipole translations; multipole-to-local transformations and translations, $c$; direct interactions, $d$; and lower order terms, $e$.

In some respects, however, the analysis in [10] was limited. All the specific cases analyzed assumed a uniform distribution of particles at the finest level. As a consequence, no strategy is offered for dealing with non-uniform distributions, and computational imbalance is not addressed. The volume of communication for a certain partition of work is also not estimated in this model. Below, we extend the model by giving estimates for both communication volume and computational load. The new, extended model will allow us to generate an unstructured data distribution which is optimal in terms of load balance and communication overhead.

### 5.1   Communication estimate

In Section 4, we discussed the parallelization strategy, where the tree representation of the FMM algorithm is decomposed into a set of subtrees. We now discuss in more detail the data communication that takes place between subtrees. Figure 3 presents a simplified illustration of the communications showing all the data usage patterns across different parts of the algorithm. The illustration helps us in the identification of three types of communications that take place between two subtrees: multipole-to-multipole translations (M2M), multipole-to-local transformations (M2L), and local-to-local translations (L2L). Figure 4 is an equivalent graph representation of the partitioned tree from Figure 3, translating graphically the hierarchical tree into a set of vertices and edges: the subtrees are represented as vertices and the communication between subtrees correspond to the edges of the graph.

In order to determine the edge weights of the graph, it is necessary to identify all the communications required between nodes of the FMM tree associated to different vertices of the equivalent graph. Communications between sub-trees at different stages of the FMM algorithm can be classified into two types: communication of particles in the local domain, and communication of *expansion terms*. This last type occurs at M2M and L2L translations, and M2L transformations. A complete picture of the communication patterns for a binary-tree can be seen in Figure 3. We note that the communication patterns for the M2M and L2L translations occur only from subtrees to root tree and vice versa, while no communication between subtrees takes place for these operations.

Quantification of the amount of communication between each pair of vertices is required. For each communication pattern, we define a communication matrix, where each element $c_{ij}$ of the matrix indicates the amount of communication from vertex $i$ to vertex $j$. We now consider the two-dimensional case, where a quadtree is used. By studying the communication patterns between nodes in the quadtree, we arrive at estimates of the communications between subtrees of the truncated tree. Thus, we produce the following estimate for the amount of communication between two lateral neighboring subtrees of a quadtree,

$$\sum_{n=k+1}^{L} \alpha_{\text{comm}} \, (2)^{n-k} * 4 \tag{11}$$

where $L$ is the depth of the FMM tree, $k$ is the cut level of the quadtree, and $\alpha_{\text{comm}}$ is a constant depending on the expansion order $p$ and the size of floating point numbers used. In the same way, we obtain the following estimate for the amount of communication between two diagonal neighboring subtrees,

$$\alpha_{\text{comm}} \, ((k - L) - 1) * 4 \tag{12}$$

since only corner boxes are involved at each level of the subtree. The last step is to construct the communication matrix for subtrees. The quadtree $z$-order numbering of the nodes is used to discover the neighbor sets for every vertex of the graph without any communication between processes. All communication occurs between neighboring local trees, so we can fill the communication matrix as follows,

```
For all node j at level k:
    For all node i from the neighbor set of j:
        if i is a lateral box:
            c[i][j] += lateral node estimate
        else
            c[i][j] += diagonal node estimate
```

### 5.2 Work estimate

In Section 2, we decomposed the FMM algorithm into three stages: upward sweep, downward sweep, and the evaluation step. In this section we discuss work estimates following the same decomposition.

In the upward sweep, the MEs for all the nodes of the FMM tree are built. It takes $O(N_i p)$ operations to build the MEs for a node that is in the maximum level of refinement (or the leaf level of the FMM tree), where $N_i$ corresponds to the number of particles associated to the box $i$, and $p$ is the number of expansion terms retained. To obtain the MEs for the upper levels of the FMM tree, each node of the upper levels *translates* into its center the MEs of their child nodes. The translation of a single ME has cost $O(p^2)$ and if a node has $n_c$ child boxes the work performed by a node is $O(n_c p^2)$.

In the downward sweep, the LEs that represent the complete far field of a leaf node are built. For each local cluster, LEs are built for all clusters in its interaction list. If the *interaction list* (IL) has $n_{IL}$ members and the cost to *transform* an ME into an LE is $O(p^2)$ then, for each node of the FMM tree, the total work to obtain the LEs for its interaction list is $O(p^2 n_{IL})$. After all the MEs are converted into LEs, the LEs are propagated from the parent box to the child boxes. For each non-leaf node of the FMM tree, the LE is translated from a node's center into its $n_c$ child boxes. The work to translate a single LE is $O(p^2)$, thus the work performed by a node is $O(n_c p^2)$.

In the evaluation step, the complete field is obtained for all the particles of the system, where the near field is computed from the local direct interactions and the far field is obtained from the LEs. The near domain is given by the leaf node that contains the particle. If, for a box, the near domain is given by $n_{nd}$ nodes, each with $N_i$ particles, the cost of computing all pairwise interactions for a leaf node is $O(n_{nd} N_i^2)$. The far field of a particle is obtained from the evaluation of the LE of the box that contains the particle. If a box has $N_i$ particles, the cost of evaluating all the far field for all the $N_i$ particles is $O(N_i p)$ where $p$ is the number of terms in the LE.

In summary, the amount of work done by a non-leaf node of the FMM tree is:

$$O(p^2(2n_c + n_{IL})) \tag{13}$$

and the amount of work performed by a leaf node is:

$$O(2N_i p + p^2 n_{IL} + n_{nd} N_i^2) \tag{14}$$

where $n_c$ corresponds to the number of children of a box, $p$ is the number of expansion coefficients retained, $n_{IL}$ is the estimate of the number of members in the interaction list, and $N_i$ is the number of particles in the box $i$.

When partitioning the FMM tree, $T$ sub-trees are created, each of which has the same number of levels $L_{st}$. The estimate for the total amount of work performed by each subtree depends on its number of nodes and its number of particles at the leaf level. By means of equations (13) and (14), we can estimate the amount of work performed by any subtree and use this as weight for the corresponding vertex of the graph:

$$\left( \sum_{l=0}^{L_{st}-2} 2^{dl} p^2 \left(2n_c + n_{IL}\right) \right) + 2^{d(L-1)} \left(2N_i p + p^2 n_{IL} + n_{nd} N_i^2\right) \tag{15}$$

Table 1
Memory usage for serial quadtree structures.

| Type | Bookkeeping Memory (bytes) | Data Memory (bytes) |
|---|---|---|
| Box centers | 0 | $8d\Lambda$ |
| Interaction boxes | $(2*4)\Lambda$ | $(27*4)\Lambda$ |
| Interaction values | $(2*4)\Lambda$ | $27(8d+16p)\Lambda$ |
| Multipole coefficients | 0 | $16p\Lambda$ |
| Temporary coefficients | 0 | $16p\Lambda$ |
| Local coefficients | 0 | $16p\Lambda$ |
| Local particles | $(2*4)\Lambda$ | $BN$ |
| Neighbor particles | $(2*4)\Lambda$ | $8Bs2^{dL}$ |

### 5.3 Memory estimate

We will first estimate memory usage in the serial case, and then extend these results to the parallel code, since most of the computational structures are preserved. The quadtree structure stores data of three types: $\mathcal{O}(1)$ storage, such as sizes, data across boxes of the finest level, and data across all nodes of the tree. Over the entire tree, we store the box centers, the interaction list and associated values, and three expansion coefficients. We store local and neighbor particles only at the finest level of the tree. Let $d$ be the space dimension, $L$ be the maximum level, $p$ the number of expansion terms, $N$ the number of particles, $B = 28$ the size of a particle in bytes, $s$ the maximum number of particles per box, and let $\Lambda$ be the total number of boxes in the tree, given by

$$\Lambda = \sum_0^L 2^{dl} = \frac{2^{d(L+1)} - 1}{3}.$$

The maximum memory usage is given in Table 1. Notice that even for very nonuniform distributions, the memory usage for neighbor particles is bounded above by a constant multiplied by the total number of particles. Thus the memory usage is linear in the number of boxes at the finest level and the number of particles.

In the parallel case, we reuse our serial data structures completely, and thus we only need to estimate the memory usage from explicitly parallel constructs. We maintain both a partition of boxes, and its inverse. We also have overlap [15] structures, essentially box-to-box maps between processes, for both neighbor particles and interaction list values. Let $P$ be the number of processes, $N_{lt}$ the maximum number of local trees, $N_{bd}$ the maximum number of boundary boxes for a process, $A = 108$ the size of an arrow in the overlap structure. The maximum memory usage per process is given in Table 2. The neighbor overlap structure is bounded by the total number of particles exchanged, whereas the interaction list overlap is bounded by the maximum cut size, a constant multiplied by $N_{lt}$.

Table 2
Memory usage for parallel quadtree structures.

| Type | Bookkeeping Memory (bytes) | Data Memory (bytes) |
|---|---|---|
| Partition | $(2*4)P$ | $4N_{lt}$ |
| Inverse partition | 0 | $4N_{lt}$ |
| Neighbor send overlap | N/A | $N_{bd}sA$ |
| Neighbor recv overlap | N/A | $N_{bd}sA$ |
| Interaction send overlap | N/A | $27N_{bd}A$ |
| Interaction recv overlap | N/A | $27N_{bd}A$ |

Future improvements may include memory savings through computing values dynamically. For instance, we need not store the interaction list boxes, as they can be generated from the global box number. Also, for the case of uniform spacing, the box centers can be determined on the fly. Neighboring particles, except for those from other processes, could be retrieved from the local particle storage. The inverse partition of boxes could be calculated on the fly, albeit with communication.

## 6 Software design, implementation and verification

### 6.1 Design of the software interfaces

The PetFMM library was designed to offer both high serial performance and scalability, but also to be easily integrated into existing codes and maintained. The serial code is completely reused in the parallel setting so that we are never required to maintain two versions of the same algorithm. PetFMM leverages existing packages to keep its own code base small and clear. Parallel data movement is handled by the *Sieve* package [15], while load and communication are balanced using ParMETIS, as described in the previous sections.

The problem geometry is encapsulated in the `Quadtree` class. All relations, such as neighbors and interaction lists, can be dynamically generated so that we need only store data across the cells. Data is stored in *Sieve* `Section` objects, which allow easy communication across unstructured boundaries with arbitrary layouts. The `ParallelQuadtree` class presents an identical interface, but manages a set of local `Quadtrees` and a root `Quadtree` to mimic these methods.

All serial computation is encapsulated in the `Evaluator` class, which operates over a `Quadtree`. It is templated over a `Kernel` object, which produces the required expansions and translation operators, so that we can easily replace one equation with another. The `ParallelEvaluator`, which inherits from `Evaluator`, overrides some methods in order to operate over a `ParallelQuadtree`. All computation in the parallel code is still done by serial structures, and only data partitioning and movement is accomplished by new code, mainly derived from the *Sieve* library.

### 6.2 Code verification

From an initial, serial Python implementation of the FMM algorithm [2], we produced a parallel C++ version in little more than two weeks (with subsequent debugging and verification over several more weeks). The PETSc *Sieve* framework [15] greatly aided parallelization. We were also careful at each stage to introduce unit testing and verify our results against those of the well-studied Python code [8]. We developed a file format for result verification, consisting of

▷ The number of levels, terms, particles, and tree coordinates
▷ Particle assignment to quadtree cells
▷ Center, number of particles, children, neighbors for each box
▷ Interaction list and coefficients for each box
▷ The direct and FMM solutions

All domain boxes were labeled with global numbers in the output, which enables the parallel code to be compared with the serial implementations. Moreover, the box output may come in any order, making parallel output trivial. A small Python script then compares two output files, noting any discrepancies. In this way we compared the initial Python code with the C++ version, both serial and parallel, and even among parallel runs with different number of processes. Confidence in our results enabled us to make rapid changes, introducing new algorithms such as the improved partitioning scheme without lengthy debugging.

---

[2]  Available for free download at http://code.google.com/p/pyfmm/

# 7 Computational experiments with the parallel software

## 7.1 Experimental setup and test case

The results presented here were obtained on the BlueCrystal–I system at the [Advanced Computing Research Centre](#), University of Bristol. The BlueCrystal–I system is composed of 416 nodes, each with two quad-core 2.8 GHz *Intel®Harpertown* processors and 8 GB in RAM. The system is interconnected with a *QLogic InfiniPath®* high-speed network.

As an experimental setup, we tested the strong scaling capabilities of PetFMM up to 64 nodes, with only one process per core in order to concentrate on network communications. The test case corresponds to the use of the FMM in the context of an application of the vortex method to a viscous flow problem; see §3. We use the Lamb-Oseen vortex, a known analytical solution of the Navier-Stokes equations, to initialize the strength (representing the amount of circulation $\gamma_i$) of each of the particles of the system, as in [4]. The core sizes of the particles are uniform for all the particles and set to $\sigma = 0.02$. The particles of the system are positioned on a lattice distribution in a square domain of fixed size; the separation between particles is given by a constant spacing parameter $h$ and is obtained from the relation $\frac{h}{\sigma} = 0.8$, as in [4].

The analytical solution for the vorticity field of the Lamb-Oseen vortex is:

$$\omega(r,t) = \frac{\Gamma_0}{4\pi\nu t} e^{-r^2/4\nu t},\tag{16}$$

where $r^2 = x^2 + y^2$, $\Gamma_0$ represents the total circulation of the vortex, $\nu$ is the viscosity of the fluid and $t$ is time. The velocity field, in turn, is given by:

$$u(r,t) = \frac{\Gamma_0}{2\pi r} \exp\left(1 - e^{-r^2/4\nu t}\right).\tag{17}$$

The analytical solution of the velocity field for the Lamb-Oseen vortex is used to compare the results obtained with solving the Biot-Savart velocity, accelerated via the FMM. Extensive studies of the accuracy of the FMM for this problem, with respect to the algorithm parameters, were reported in [8]. Many of the experiments reported there were repeated with PetFMM, as part of the verification stage of the parallel program. For the strong scaling experiments reported below, we designed a problem setup with parameters which were appropriate to test the parallel performance and strain communications, but which may not be optimal from the point of view of accuracy of the computed field. In particular, the scaling study used many levels in the tree, which introduces errors of Type I, as reported in [8], related to kernel substitution.

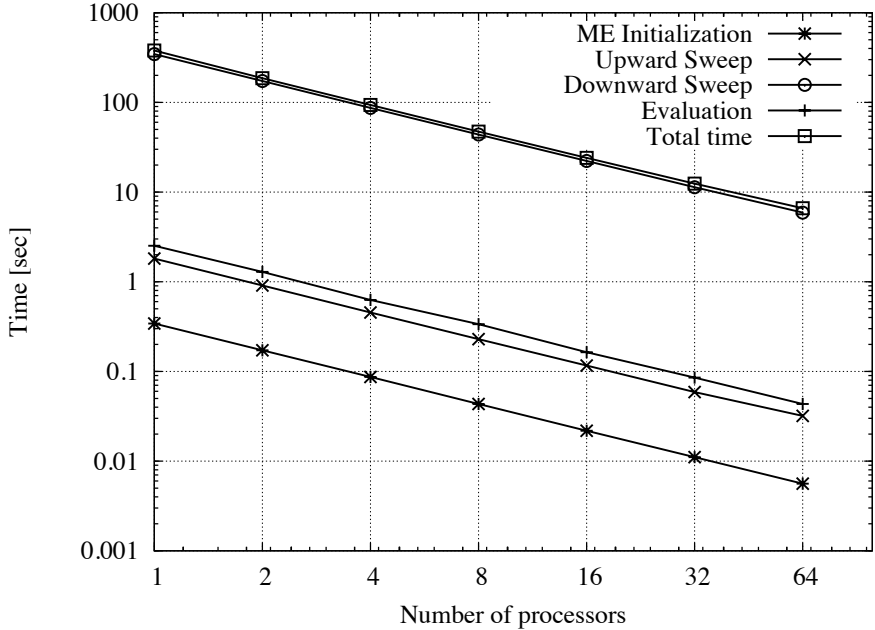

Fig. 6. Measured time for experimental results obtained with PetFMM, presented in base 10 logarithm scale, for an increasing number of processors, presented in base 2 logarithm scale. The timings for the total execution of PetFMM (labeled as *Total time*) and for the the more important stages of the algorithm are presented.

### 7.2 *Experimental results*

We now present the results of numerical experiments with PetFMM. In order to present the scalability of PetFMM, we report strong scaling results, *i.e.*, we compare how the total execution time of PetFMM varies with the number of processors for a fixed problem size. The same set of parameters were used for all the experimental results: $N = 765,625$, level 10, root level 4, and $p = 17$. Therefore, the total work remains the same but we vary the number of processors used: $P = \{1, 4, 8, 16, 32, 64\}$. Figure 6 shows the measured time for experimental results obtained with PetFMM for increasing number of processors. Measured times for the more important stages of the FMM algorithm are also given, effectively depicting the fraction of time consumed by each stage of the algorithm.

For the analysis of performance for PetFMM, we present the speedup, parallel efficiency and a load balancing metric. For a definition of speedup $(S)$ we use

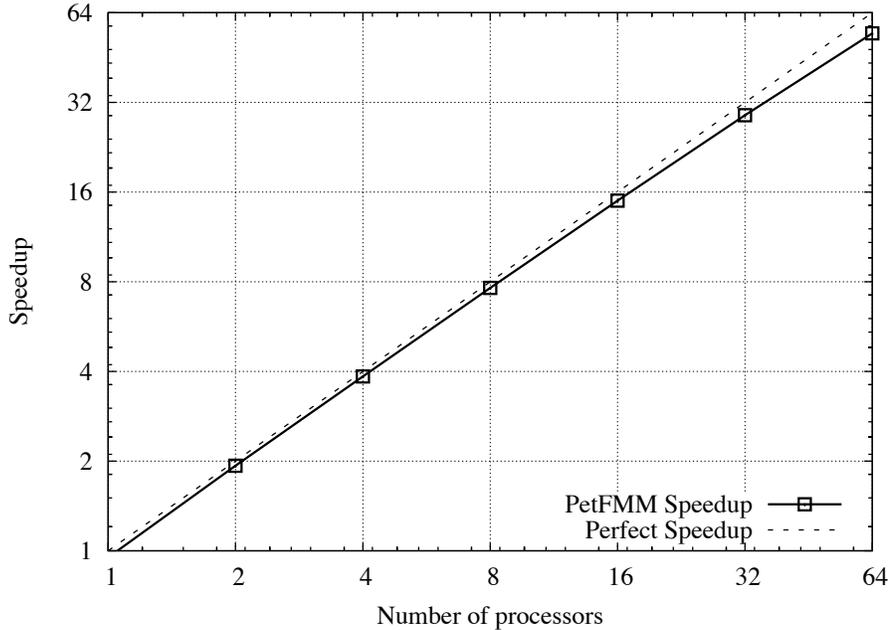

Fig. 7. Speedup of PetFMM for an increasing number of processors. The speedup and the number of processors are presented in a base 2 logarithm scale. The deviation of PetFMM from perfect speedup indicates that there are still some serial stages within the algorithm and communication overheads.

the factor given by

$$S(N, P) = \frac{\text{execution time for serial case}}{\text{execution time for } P \text{ processors}} \qquad (18)$$

A perfect speedup means that the parallel program perfectly scales with the number of processors. This is almost never achieved, due to the existence of serial stages within an algorithm and communication overheads of the parallel implementation. Figure 7 shows the speedup of PetFMM up to 64 processors, and compares it against a perfect speedup. Despite the existence of serial stages, the speedup achieved is excellent for a first version of the software. Scaling to larger numbers of processors will require additional algorithm improvements; some thoughts about how we expect to achieve this are given in the Conclusions.

The definition of parallel efficiency ($E$) that we use is:

$$E(N, P) = \frac{S(N, P)}{P}. \qquad (19)$$

Parallel efficiency equal to unity means that the parallel implementation scales perfectly with the serial implementation; this is equivalent to perfect speedup, but efficiency is a more revealing metric. Both of these performance metrics assume that the instructions executed by the parallel application are the same as the instructions executed by the serial application. Of course, this is not the case. In general, the differences between serial and parallel implementations are due to: different workloads between the serial and parallel processes, and the existence in the parallel implementation of data communication among processes and synchronization of processes. As a consequence of these differences, perfect speedup or perfect efficiency ($E = 1$) are generally not obtained. In fact, as seen in Figure 8, it can happen that an efficiency value greater than unity is obtained, which reveals the shortcomings of the metric. Figure 8 shows the parallel efficiency of PetFMM for the main stages of the FMM algorithm.

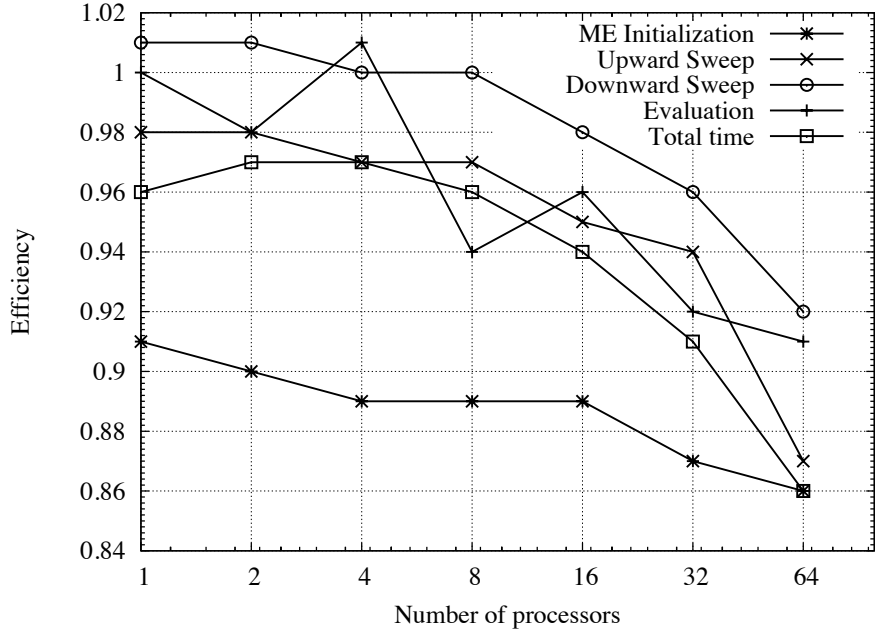

Fig. 8. Parallel efficiency of PetFMM for an increasing number of processors. Here only the relevant range of efficiency is plotted against the number of processors in a base 2 logarithm scale.

Figure 9 shows a load balance metric, which we define as

$$LB(P) = \frac{\text{minimum execution time for one of } P \text{ processors}}{\text{maximum execution time for one of } P \text{ processors}} \quad (20)$$

A load balance metric of unity would indicate equal execution times among all processors in a parallel run, and thus perfect load balancing. Figure 9 also includes the total parallel efficiency, which follows a similar trend, indicating that parallel overheads and communications overheads are correlated with a degradation of the load balance metric. Nevertheless, it can be seen that processor execution times are within 5% of each other for 32 processors and within 7% of each other for 64 processors, demonstrating the success of our load balancing strategy.

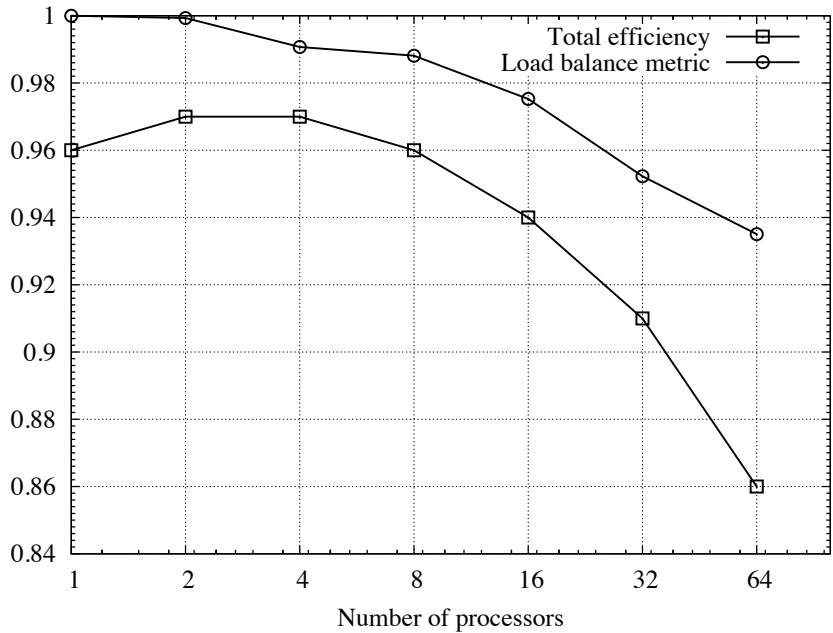

Fig. 9. Load balance metric of PetFMM for an increasing number of processors. Here only the relevant range of efficiency is plotted against the number of processors in a base 2 logarithm scale. The load balance metric is compared with the total efficiency of the runs.

The largest problem size that has been run with PetFMM at this point solved for 64 million particles on 64 processors. The total execution time was 115.8 seconds and the amount of memory used was under 1.01 GB per processor. PetFMM is currently able to achieve over 90% parallel efficiency for 32 processors, and over 85% parallel efficiency for 64 processors.

## 8 Conclusions

We have presented an extension of the Greengard-Gropp parallel complexity analysis for the FMM which is suitable for distributed computing. We thus consider partition quality in terms of both communication volume and load balance, absent from the original analysis, in order to derive an optimally distributed algorithm. This algorithm can maintain good performance with highly non-uniform particle distributions, heterogeneous computing environments, and low bandwidth connections.

We have also delivered an open source, extensible implementation of this algorithm, intended as a community resource. The software library is truly open source[3], and distributed under the PETSc license, even less restrictive than the BSD license. This guarantees the maximum impact to the scientific community and encourages peer-based collaboration for the extensions and applications.

Our original approach to parallelization relies on recasting the hierarchical space division represented by a tree structure, into a weighted graph generated from a model of the computational work and communications estimates. The optimal partitioning of the graph produces dynamical load balancing. Experiments showed that processor execution times were within 5% of each other for 32 processors and within 7% of each other for 64 processors, demonstrating the success of our load balancing strategy. The parallelization strategy permitted almost complete reuse of the serial implementation, and due to its simplicity, was effected with very little additional code. It has shown very good strong scaling on small clusters. The library is currently able to achieve over 90% parallel efficiency for 32 processors, and over 85% parallel efficiency for 64 processors. The largest problem size that has been run with the PetFMM library at this point was 64 million particles in 64 processors.

At present, we are working on exposing more concurrency in the algorithm to exploit multi-core architecture; our real goal here is to extend these results to large clusters, and also to heterogeneous systems, including programmable graphics processing units, GPU. With respect to delivering scalability to larger numbers of processing, we are investigating a strategy based on recursive tree-cutting approach, where the methodology described here can generate large subtrees which are further cut to obtain sub-subtrees.

In terms of the extensibility of PetFMM, the first and most straightforward extension is to introduce 3D capability. Our methodology applies without modification to 3D, as our tree and particle objects are templated over the dimension. Moreover, the parallel model involving work and communications estimates, and the setup for optimization of the partition, all apply to 3D. The main modification here involves the implementation of a new kernel, incorporating the 3D series expansions, and tranlation/transformation operators.

Future extensions of PetFMM include the implementation of different kernels, for example Lennard-Jones potential for molecular dynamics applications, and extensions to other transforms such as fast Gauss transform. We expect more extensions can be provided in the future by peers who find PetFMM to be a good starting point for their scientific applications. To encourage peer-based collaboration, we are working to deliver an official release by the end of the year, consisting of a User's Manual, example code and verification methods.

---

[3] According to the definition provided in [http://www.opensource.org/docs/definition.php](http://www.opensource.org/docs/definition.php)

## Acknowledgments

Computing time provided by the Advanced Computing Research Centre, University of Bristol. FAC acknowledges financial support from Airbus and BAE Systems under contract ACAD 01478. LAB acknowledges partial support from EPSRC under grant contract EP/E033083/1, and from Boston University College of Engineering.